\documentclass[biblatex]{biophys-new}
\usepackage{graphicx,booktabs,amssymb}
\usepackage[colorlinks=true,allcolors=blue!45!black]{hyperref}
\usepackage{placeins}
\newcommand{\dk}{\Delta_\kappa}
\newcommand{\dm}{\Delta_{\kappa m}}
\newcommand{\dg}{\Delta_{\bar\kappa}}
\newcommand{\tk}{\tilde{\Delta}_\kappa}
\newcommand{\tg}{\tilde{\Delta}_{\bar\kappa}}
\newcommand{\hm}{\hat{\Delta}_{\kappa m}}
\newcommand{\hg}{\hat{\Delta}_{\bar\kappa}}

\title{Curvature-induced migration of small domains on lipid membranes}
\author[1]{Jaime Agudo-Canalejo}
\affil[1]{Department of Physics and Astronomy, University College London, London WC1E 6BT, United Kingdom\\j.agudo-canalejo@ucl.ac.uk}
\runningtitle{Curvature-induced domain migration}
\runningauthor{Agudo-Canalejo}
\papertype{Research Article --- Draft}

\makeatletter
\patchcmd{\@maketitle}{Manuscript submitted to}{Manuscript prepared for}{}{}
\makeatother
\begin{document}
\begin{frontmatter}
\begin{abstract}
Membrane curvature has been shown to bias the location of lipid or protein domains, but the roles of bending rigidity, spontaneous curvature, Gaussian bending modulus, and line tension are difficult to separate in general. We develop a small-domain description for a membrane whose shape is held fixed, e.g. by strong adhesion to a curved substrate. Combining a local Helfrich energy with the small-area isoperimetric expansion gives a position-dependent energy determined by the local mean and Gaussian curvatures. We identify four generic regimes of curvature preference whose boundaries depend on only two dimensionless parameters. On specific membrane shapes with complex curvature, we show that this curvature preference determines the energy landscapes experienced by small domains, which are strongly dependent on the size of the domain through the line tension contribution. For example,  on an oblate-shaped membrane, very small domains tend to localize at the poles, whereas larger domains prefer to localize at the equator. Whether this transition is continuous or discontinuous is found to strongly depend on the spontaneous curvature of the domain. This ordering suggests a mechanism by which small domains can collect and coalesce at one location before relocating as they grow.
\end{abstract}

\begin{sigstatement}
A membrane's shape can provide a spatial cue without a chemical pattern. Here we derive a simple energy landscape for a small lipid or protein domain on a curved membrane. A key result is that domain size changes the contribution of line tension relative to bending elasticity, allowing the preferred location to switch as domains grow. Calculated diagrams for simple prolate and oblate spheroids show especially clearly how this mechanism depends on spontaneous curvature and geometry, and when the migration is continuous or discontinuous. The framework provides testable predictions for mobile domains on curved supports, which could be applicable to biological membranes.
\end{sigstatement}
\end{frontmatter}

\section{Introduction}

Lipid composition and membrane shape are coupled aspects of membrane organization. Coexisting fluid domains can be directly observed in model membranes \cite{veatch2004,baumgart2003imaging}, and their boundaries, bending properties, and local curvature influence one another \cite{julicher1996shape,hu2011vesicles}. Curvature-dependent sorting has been measured in membrane tubes and in bilayers on patterned supports \cite{parthasarathy2006,tian2009,sorre2009}. More recently, scaffolded vesicles have provided a way to impose a complex shape while monitoring the distribution of lipid phases \cite{rinaldin2020}. Experiments and simulations on nanoscale supported buds further show that phase behavior, lipid sorting, and diffusion must often be considered together \cite{woodward2023}. These observations motivate a basic question: given an existing lipid domain and a membrane shape, where does the domain prefer to go?

Several related problems have been explored in the literature. One recent line of work considers the equilibrium interfaces between macroscopic membrane phases. For a prescribed curved membrane, the equilibrium boundary depends on line tension, bending and Gaussian moduli, spontaneous curvature, and the area assigned to each phase \cite{fonda2018interface,fonda2019thermodynamic}. Mathematically, this question is related to the isoperimetric problem, which asks how global geometry restricts the shape and location of perimeter-minimizing regions \cite{morgan2000isoperimetric,morgan2000some}. A related problem is that of nucleation on a curved surface. Here, the cost of a phase boundary competes with the free-energy gain from creating the new phase. Curvature changes the relationship between enclosed area and perimeter and therefore changes nucleation energetics \cite{gomez2015phase}. On nonuniform surfaces, the preferred phase and its preferred location can become coupled, as has been shown using computer simulations \cite{law2020phase}. Importantly, nucleation of a domain occurs stochastically as a result of thermal fluctuations, and might in principle happen anywhere on a curved surface. Which material parameters of a domain couple to which local features of a curved membrane to direct the migration of a domain remains an open question, which we tackle here by developing an analytical theory in the limit of small domains.

Previously, this approach has been used successfully to understand the curvature-induced migration of solid particles adhering to membranes. For sufficiently small adhering particles, a position-dependent adhesion and deformation energy can be expressed in terms of local membrane curvature, both in the case of tensionless membranes dominated by bending \cite{agudo2017uniform}, and of tense membranes dominated by capillarity \cite{li2017curvature}. The resulting energy gradients bias particle migration. In the former case, it was found that particles coming from the outside migrate toward regions of lower mean curvature (average of principal curvatures), whereas in the latter they do so toward larger absolute deviatoric curvature (difference between principal curvatures).

\begin{figure}[tbp]
\centering\includegraphics[width=0.6\linewidth]{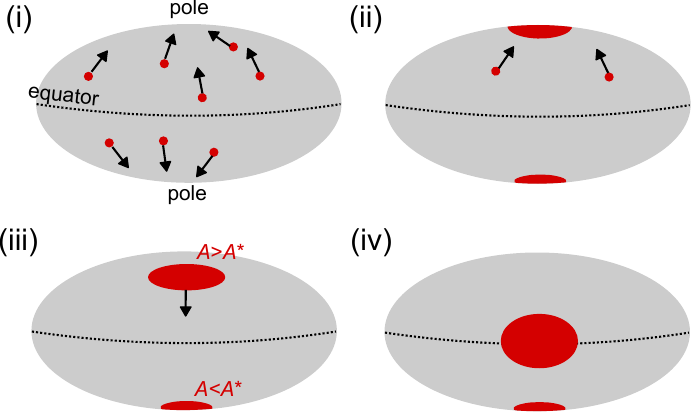}
\caption{Schematic of a typical sequence (i--iv) on an oblate-shaped membrane, showing how domain growth can change the preferred location on a curved membrane. (i,ii) Small domains are initially attracted to the poles, where they collect and fuse. (iii,iv) If a domain grows beyond a critical size $A_*$ (loss of stability threshold), it will spontaneously migrate toward the equator, otherwise it will remain at the pole.}
\label{fig:overview}
\end{figure}

We study the curvature-induced migration of small membrane domains on membranes with fixed shape, representing e.g.~strong adhesion to a curved substrate which, in the biological case, could also represent a cell wall or cortex. To describe the energetics of the domains, we combine the Helfrich description of their membrane elasticity \cite{canham1970,helfrich1973}  with the small-area isoperimetric expansion of Nardulli \cite{nardulli2009} to account for the line tension contribution. The resulting energy contains only a term proportional to the local mean curvature squared, governed by the bending rigidity contrast between the domain and the surrounding membrane; a term linear in the local mean curvature, governed by the rigidity-weighted spontaneous curvature contrast; and a term linear in the local Gaussian curvature, governed by the Gaussian modulus contrast and a domain-size-dependent line tension contribution. From this energy, we first derive a universal classification of curvature preference by small domains that depends only on two dimensionless parameters. We then study domain migration on specific membrane shapes, in particular on prolate and oblate ellipsoids, and show how material parameters determine the preferred location of domains on these shapes. Importantly, increasing domain size affects the coupling to the local Gaussian curvature, where larger domains prefer larger Gaussian curvature. Figure~\ref{fig:overview} illustrates one consequence of this effect: on an oblate membrane shape, small domains first migrate to the poles. After growth through coalescence beyond a critical size, they then subsequently migrate to the equator.

\section{Methods}
\subsection{A small domain on a membrane with fixed shape}
We consider a smooth membrane surface and an isolated fluid domain of area $A$. Domain and background have bending rigidities $\kappa_d$ and $\kappa$, Gaussian moduli $\bar\kappa_d$ and $\bar\kappa$, and spontaneous curvatures $m_d$ and $m$. Their interface has a positive line tension $\lambda$. We use the convention
\begin{equation}
M=\frac{k_1+k_2}{2},\qquad K=k_1k_2,
\label{eq:curvatures}
\end{equation}
where the  principal curvatures $k_1$ and $k_2$ are positive on a sphere.
Strong adhesion to an external scaffold is assumed to keep $M$ and $K$ fixed as the domain moves. The energy, defined relative to that of the background membrane in the absence of the domain, is
\begin{equation}
E=\int_A [2\kappa_d(M-m_d)^2-2\kappa(M-m)^2
 +\bar\kappa_d K -\bar\kappa K]\,dA+\lambda P
\label{eq:fullenergy}
\end{equation}
where $P$ is the perimeter of the domain, which is a topological disk with a narrow interface compared with its radius.

\begin{figure}[tbp]
\centering\includegraphics[width=\linewidth]{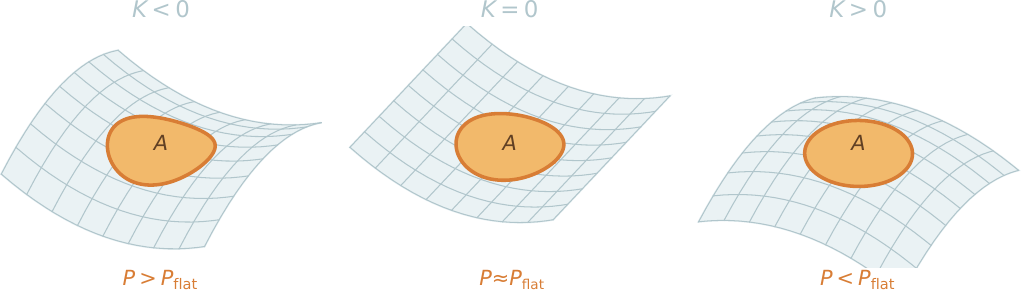}
\caption{Curvature-dependence of the line-tension contribution. Small domains of equal area $A$ on schematic saddle-like ($K<0$), developable/flat ($K=0$), and dome-like ($K>0$) regions. The first correction to $P_{\rm flat}=2\sqrt{\pi A}$ increases the perimeter for negative Gaussian curvature and decreases it for positive Gaussian curvature. The middle region is curved in space but intrinsically flat.}
\label{fig:perimeter}
\end{figure}

We now take the limit of a small domain. Define the area-equivalent radius $r\equiv \sqrt{A/\pi}$. The small-domain limit requires $r|k_1|\ll1$ and $r|k_2|\ll1$, together with slow variation of the curvatures across the domain. For such a small domain with area $A$ centered at a point with curvature $K$, Nardulli's local isoperimetric expansion (specialized to two dimensions) gives the leading curvature correction to the minimal perimeter as  \cite{nardulli2009}
\begin{equation}
P(K,A)=2\sqrt{\pi A}\left[1-\frac{KA}{8\pi}+O(A^2)\right].
\label{eq:perimeter}
\end{equation}
the remainder depends on the
local curvature and its spatial derivatives, Note that a geodesic disk has the same displayed correction, but an exactly perimeter-minimizing boundary need not be a geodesic circle on a variable-curvature surface. As expected, at a flat point with $K=0$ we recover the perimeter of a flat disk, $P_\mathrm{flat}=2\sqrt{\pi A}=2\pi r$. Equation~\ref{eq:perimeter} has a direct physical interpretation (Fig.~\ref{fig:perimeter}). At the same small area, a domain needs a longer boundary on a negatively curved region (saddle) and a shorter boundary on a positively curved region than in a flat plane. Positive line tension therefore favors larger Gaussian curvature when the bulk elastic contributions are held equal. We note as well that, within the domain of applicability of Eq.~\ref{eq:perimeter}, with $|K|A\ll 1$, coalescence of two domains to form a larger one always reduces the total perimeter, i.e.~$P(K,2A)-2P(K,A)<0$, implying that coalescence is favored independently of the local Gaussian curvature $K$.

Additionally, in the limit of a small domain the mean and Gaussian curvatures are approximately constant over the domain, and the integral in Eq.~\ref{eq:fullenergy} becomes trivial, with the next order of corrections being quadratic in $A$. Using this result and Eq.~\ref{eq:perimeter} then gives our central result, asymptotically valid in the limit of small $A$:
\begin{equation}
E(M,K) = A\left[2\dk M^2
-4 \dm M
+\left(\dg-\frac{\lambda r}{4}\right)K\right]
\label{eq:central}
\end{equation}
where $\dk \equiv \kappa_d-\kappa$, $\dm \equiv \kappa_d m_d-\kappa m$,  and $\dg \equiv \bar\kappa_d-\bar\kappa$ are the bending rigidity contrast, rigidity-weighted spontaneous curvature contrast, and Gaussian modulus contrast, respectively. Terms independent of the local curvature have been omitted, as they only contribute a baseline to the curvature-dependent energy landscape. Because the energy in Eq.~\ref{eq:central} depends on the local curvature of the membrane, the energy depends on the position of the domain along a membrane with non-constant curvature, resulting in domain migration toward points that are local minima of this energy. The mechanical force experienced by the domain will be $-\nabla E$, where $\nabla$ is to be interpreted as the gradient operator in curvilinear coordinates. On an axisymmetric surface, this force has the direction of $-dE/ds$, where $s$ is the arclength (distance along the meridian).

\subsection{Geometry of axisymmetric spheroids}

In Section~\ref{sec:shapes}, we will illustrate our results using axisymmetric spheroids. These can be parameterized as
\begin{equation}
\mathbf X(u,\phi)=
\left(
a\sin u\cos\phi,\,
a\sin u\sin\phi,\,
c\cos u
\right),
\qquad
0\le u\le\pi,\quad 0\le\phi<2\pi,
\end{equation}
where $a$ and $c$ are the equatorial and polar semi-axes,
respectively. Defining the function
\begin{equation}
Q(u) = \sqrt{a^2\cos^2u+c^2\sin^2u}
\end{equation}
for conciseness, the principal curvatures can be written as
\begin{equation}
k_1(u)=\frac{ac}{Q(u)^3},
\qquad
k_2(u)=\frac{c}{aQ(u)}.
\end{equation}
Consequently, the mean and Gaussian curvatures are given by
\begin{equation}
M(u)=
\frac{c\left[a^2+Q(u)^2\right]}
{2aQ(u)^3},
\qquad
K(u)=\frac{c^2}{Q(u)^4}.
\end{equation}
In particular, at the poles the two curvatures are $M_\mathrm{p}=c/a^2$ and $K_\mathrm{p}=c^2/a^4$, while at the equator they are $M_\mathrm{e}=(a/c^2+1/a)/2$ and $K_\mathrm{e}=1/c^2$.

The total surface area $S$ of the spheroid can be calculated as
\begin{equation}
S=
2\pi a\int_0^\pi
\sin u\,
\sqrt{a^2\cos^2u+c^2\sin^2u}\,du.
\end{equation}
Defining the aspect ratio $q \equiv c/a$, we find that for a prolate spheroid with $q>1$, the area is given by
\begin{equation}
S = 2 \pi a^2 \left( 1 + \frac{q^2}{\sqrt{q^2-1}} \mathrm{arcsin}\sqrt{1-\frac{1}{q^2}} \right)
\end{equation}
and, for an oblate spheroid with $q<1$, by
\begin{equation}
S = 2 \pi a^2 \left( 1 + \frac{q^2}{\sqrt{1-q^2}} \mathrm{arctanh}\sqrt{1-q^2} \right).
\end{equation}

For comparison between shapes, we define the area-equivalent radius $L\equiv \sqrt{\frac{S}{4\pi}}$ which will be used below. We will use one prolate example with aspect ratio $q=1.25$ and one oblate example with $q=0.8$.
For spheroids, the mean and Gaussian curvatures vary monotonically with latitude. For the prolate spheroid, both $M$ and $K$ are lowest at the equator and increase toward the poles, while the opposite is true for the oblate spheroid, with $M$ and $K$ largest at the equator and decreasing toward the poles.

\section{Results}
\subsection{A generic classification of curvature preference}
\subsubsection{Constraints between mean and Gaussian curvature}

We will first investigate the generic consequences of Eq.~\ref{eq:central}, i.e.~those that are independent of a particular choice of surface. Equivalently, if a surface presented all possible combinations of mean and Gaussian curvature $M$ and $K$, where would a small domain most prefer to sit? Importantly, mean and Gaussian curvature cannot be varied completely independently. From Eq.~\ref{eq:curvatures}, one finds that 
\begin{equation}
M^2-K=\frac{(k_1-k_2)^2}{4} \ge 0.
\label{eq:constraint}
\end{equation}
We define the deviatoric curvature $D \equiv (k_1-k_2)/2$, such that $D^2$ measures how different the two principal curvatures are, and which satisfies $M^2-K=D^2$. Importantly, $M$ and $D$ can be varied independently of each other. Using these two curvatures,  Eq.~\ref{eq:central} becomes
\begin{equation}
E(M,D) = A\left[ \left( 2\dk + \dg-\frac{\lambda r}{4} \right) M^2
-4 \dm M
+\left(\frac{\lambda r}{4} - \dg\right)D^2\right]
\label{eq:deviatoricfull}
\end{equation}
This form of the energy clearly separates a mean-curvature mode from a curvature-difference mode, enabling a classification. Note that a point with equal principal curvatures, i.e.~$D=0$ or equivalently $M^2=K$, corresponds to an umbilic point. Such points are locally spherical to quadratic order when $M\ne0$ and flat to quadratic order when $M=0$.

\subsubsection{Four generic regimes of curvature preference}

The signs of the two quadratic coefficients in Eq.~\ref{eq:deviatoricfull} define four regimes with regard to the curvature preference of the small domain.

Regime I corresponds to both coefficients being positive, which implies
\begin{equation}
\frac{\lambda r}{4} - 2 \dk < \dg<\frac{\lambda r}{4}
\label{eq:wedge}
\end{equation}
for which a necessary condition is that $\dk>0$, i.e.~that the domain be more rigid than the surrounding membrane. In this regime, the energy is minimized at umbilic points with $D=0$ and at a specific mean curvature $M=M_*$ given by
\begin{equation}
M_* \equiv 
\frac{2\dm}{2\dk+\dg-\frac{\lambda r}{4}}
\label{eq:target}
\end{equation}
whose sign is governed by the sign of $\dm$. If $\dm=0$, the domain prefers to sit on a locally flat region.

Regime II corresponds to the $M^2$ coefficient being negative while the $D^2$ coefficient is positive. This can always be achieved for a sufficiently large domain satisfying
\begin{equation}
\frac{\lambda r}{4} > \mathrm{max}\left(\dg,~\dg+2\dk\right).
\label{eq:critdomain}
\end{equation}
In this regime, deviatoric curvature is penalized, meaning that the domain will seek umbilic points, but the squared mean curvature contribution is unbounded from below. Therefore, the domain will seek the most curved umbilic points, with the sign of the mean curvature energetically biased to prefer the same sign as $\Delta_{\kappa m}$ when  $\Delta_{\kappa m}\neq0$.

\begin{figure}[tbp]
\centering\includegraphics[width=0.5\linewidth]{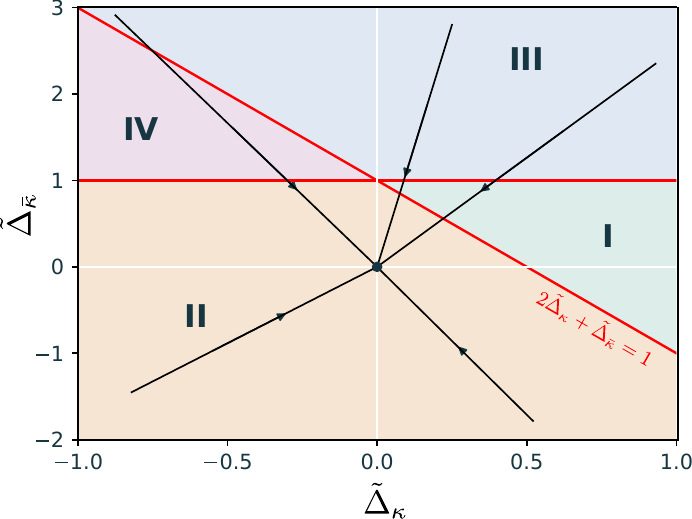}
\caption{Universal diagram of curvature preferences for a small domain, as a function of the dimensionless parameters $\tk \equiv 4{\dk}/({\lambda r})$ and $\tg \equiv 4{\dg}/({\lambda r})$. The red lines $\tg=1$ and $2\tk+\tg=1$ divide the plane into four generic regions. Region I favors umbilic points with finite mean curvature given by Eq.~\ref{eq:target}. Region II favors umbilic points with large mean curvature magnitude. Region III favors increasing curvature difference with finite mean curvature.  Region IV favors both increasing mean curvature magnitude and increasing curvature difference. Inward black arrows show the flow of the dimensionless parameters as the domain size grows at fixed material parameters, from five arbitrary starting points.}
\label{fig:universal}
\end{figure}

Regime III corresponds to the opposite case, with positive $M^2$ coefficient and negative $D^2$ coefficient. This regime occurs for sufficiently large Gaussian modulus contrast satisfying
\begin{equation}
\dg >  \mathrm{max}\left(\frac{\lambda r}{4},~\frac{\lambda r}{4}-2\dk\right).
\label{eq:regime3}
\end{equation}
In this regime, the domain will seek a mean curvature $M=M_*$ (as in Eq.~\ref{eq:target}), but will also seek to maximize the deviatoric curvature (i.e.~will seek saddle points), as the corresponding contribution to the energy is unbounded from below.

Finally, Regime IV corresponds to both coefficients being negative, which implies
\begin{equation}
\frac{\lambda r}{4}  < \dg<\frac{\lambda r}{4} - 2 \dk
\label{eq:regime4}
\end{equation}
for which a necessary condition is that $\dk<0$, i.e.~the domain is softer than the surrounding membrane. In this regime, both the mean curvature and deviatoric curvature contributions to the energy are unbounded from below, and the domain will seek extreme points with both large mean curvature and large deviatoric curvature.

Defining the two dimensionless parameters
\begin{equation}
\tk \equiv \frac{4\dk}{\lambda  r},~~~
\tg \equiv \frac{4\dg}{\lambda  r},
\label{eq:universalparameters}
\end{equation}
which control the strength of the bending rigidity and Gaussian modulus contrasts relative to the line tension contribution, these four regimes can be displayed in a universal curvature preference diagram, shown in Fig.~\ref{fig:universal}. Importantly, the values of these two dimensionless parameters tend toward zero as the domain grows, directing the system toward Regime II (provided the small-domain approximation remains valid). This reflects the increasing relative importance of the curvature-dependent line-tension contribution, which favors larger Gaussian curvature among the locations available on the membrane.

\subsection{Localization diagrams on two concrete membrane shapes \label{sec:shapes}}

The universal diagram classifies which combinations of mean and Gaussian curvatures (or, equivalently, mean and deviatoric curvature) are energetically preferred when the local curvatures are treated as independent variables subject only to the geometric constraint $K\le M^2$. A concrete membrane, however, samples only the one-dimensional (if axisymmetric) or two-dimensional subset of $(M,K)$ values actually realized on that surface. The universal optimum (regime I) may therefore be absent, while a direction where energies are formally unbounded from below (regimes II, III, IV) can be cut off by the finite range of accessible curvatures. Consequently, the universal regimes above should be interpreted as local curvature preferences rather than direct predictions of domain position on a concrete membrane shape: the actual localization pattern is obtained only after restricting the energy to the curvature landscape of the chosen membrane shape. We investigate this below, for a prolate and an oblate spheroid.

\begin{figure}[tbp]
\centering\includegraphics[width=\linewidth]{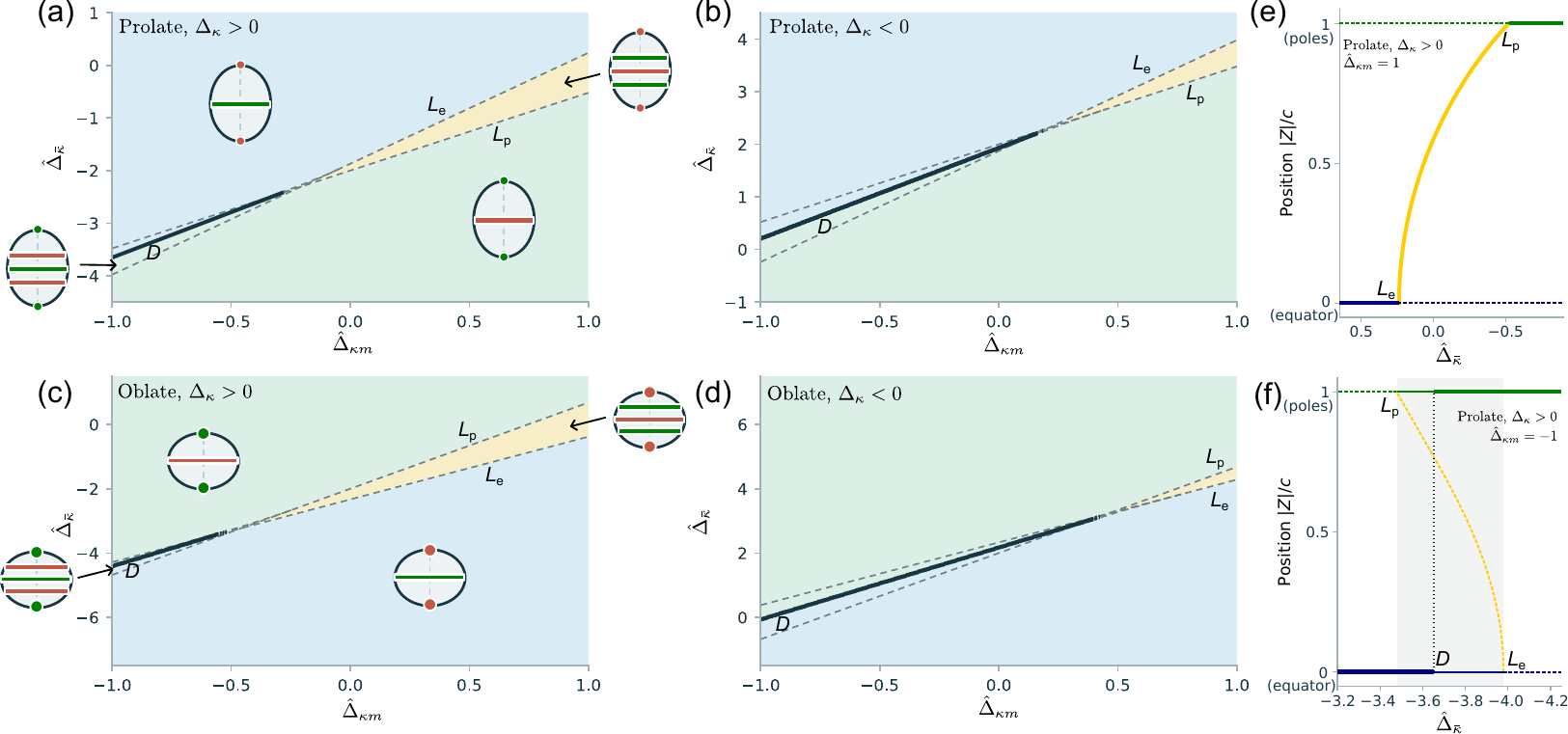}
\caption{Curvature-dependent localization on spheroidal shapes, as a function of the two material parameters in Eq.~\ref{eq:shapeparameters}. Calculated localization diagrams for a prolate spheroid with aspect ratio $q=1.25$ (a,b) and an oblate spheroid with $q=0.8$ (c,d), with domains stiffer (a,c) and softer (b,d) than the background. Domain growth corresponds to lowering $\hg$ at constant $\hm$ (i.e. downwards trajectory on the diagram). Color of the different regions gives the position of the global energy minimum: equator (blue), poles (green), or  intermediate latitude (yellow). Dashed lines correspond to changes of local stability at the poles ($L_\mathrm{p}$) and equator ($L_\mathrm{e}$), the solid line to the discontinuous transition ($D$) at which competing minima have equal energy. The cartoons in (a,c) denote the fixed points (or lines, due to axisymmetry) on the spheroid. Stable fixed points (local energy minima) are in dark green, unstable fixed points (local energy maxima) in dark red. (e,f) Bifurcation diagrams as a function of decreasing $\hg$ (equivalent to domain growth) for the prolate shape and stiff domains $\dk>0$, and two values of the spontaneous curvature contrast (e) $\hm=1$ and (f) $\hm=-1$, demonstrating a continuous and discontinuous transition from equator to pole, respectively. Dashed lines are unstable branches, solid lines stable branches. In (f), thick solid lines represent the lowest energy branch, which switches between equator and poles at the point $D$.   }
\label{fig:shapes}
\end{figure}

In the following, we choose the area-equivalent radius $L \equiv \sqrt{S/(4\pi)}$ of the spheroid as the characteristic length scale. For a finite bending rigidity contrast $\dk\neq0$, we can define two dimensionless parameters
\begin{equation}
\hm \equiv \frac{L\dm}{|\dk|},
\qquad
\hg \equiv \frac{\dg-\lambda r/4}{|\dk|},
\label{eq:shapeparameters}
\end{equation}
together with dimensionless curvatures $\hat{M} \equiv ML$ and $\hat{K} \equiv KL^2$. The dimensionless energy landscape $\hat{E} \equiv L^2 E/(|\dk|A)$ as a function of the curvatures then depends only on $\hm$ and $\hg$, as well as on the sign of $\dk$, via
\begin{equation}
\hat{E}
=2\,\mathrm{sgn}(\dk) \hat{M}^2-4 \hm \hat{M}+ \hg \hat{K}.
\label{eq:shapeenergy}
\end{equation}
The factor $L^2$ is needed because $E/(|\dk|A)$ alone is not dimensionless. For each chosen shape, two planar phase diagrams (one for each sign of $\dk$) therefore describe all possible energy landscapes experienced by a small domain (Fig.~\ref{fig:shapes}). Importantly, we note that $\hg$  depends on the domain size via the line tension contribution, and in particular decreases toward more negative values with increasing domain size. Thus, trajectories of increasing domain size correspond to decreasing $\hg$ at constant $\hm$.

For the prolate spheroid, both $\hat{M}$ and $\hat{K}$ increase monotonically from the equator to the poles. Because growth lowers $\hg$, we would expect that the preferred location of the domain should progress toward the poles as it grows. Figure~\ref{fig:shapes}(a,b) shows that this is indeed the case. Large enough $\hg$ selects the equator, while sufficiently small $\hg$ selects the poles. The opposite is true for the oblate spheroid, for which $\hat{M}$ and $\hat{K}$ are largest at the equator and decrease toward the poles. Here we find that, as $\hg$ decreases during domain growth, the preferred location changes from the poles toward the equator, as seen in Figure~\ref{fig:shapes}(c,d).

Interestingly, in both cases the transition can be continuous or discontinuous. Taking the prolate spheroid as an example, in a continuous transition, the domain moves progressively from the equator to one of the two poles as it grows, as the energy minimum progressively shifts from the equator to the poles.  In a discontinuous transition, local minima at the poles and at the equator coexist. A growing domain may remain at the equator even after this minimum has become a metastable, higher energy state relative to the lower energy minimum at the pole. In the absence of a thermally-activated escape, only beyond a critical domain size does the local minimum at the equator lose stability and the domain migrates directly to one of the two poles. Bifurcation diagrams for both the continuous and discontinuous case are shown in Figure~\ref{fig:shapes}(e,f).

In both cases, we find that whether the transition is continuous or discontinuous depends primarily on the spontaneous curvature contrast $\hm$. Above a critical value of $\hm$, the transition is continuous, while below this critical value the transition is discontinuous. Note that very close to the critical value, near the boundary between these behaviors, a discontinuous jump involving an intermediate latitude can be followed or preceded by continuous migration.
The critical value of $\hm$ is negative for $\dk>0$ (domain stiffer than background), and positive for $\dk<0$ (domain softer than background). For the particular case of zero spontaneous curvature for both domain and background (which corresponds to $\dm=0$), this implies that stiff domains undergo a continuous transition whereas soft domains undergo a discontinuous one.

Taken together, the two spheroid examples show that the small-domain theory developed here does not require geometrically elaborate shapes to produce rich behavior. Smooth, monotonic curvature profiles are already enough to realize both continuous and discontinuous migration. On more complex shapes, coexistence between multiple local minima, as well as a multitude of bifurcations and thus domain migration between these minima can be expected.

\section{Discussion and conclusion}

\subsection{Quantitative estimates and experimental verification}

In order to estimate whether the effects predicted here are observable in experiment, be it with model membranes or in cells, it is important to place bounds on the key parameters and to estimate their experimental values. Rigorous bounds can be put on the Gaussian modulus contrast $\dg$ based on general considerations. In particular, local nonnegativity of the Helfrich energy density \cite{helfrich1973} with respect to arbitrary principal curvatures requires \cite{deserno2015fluid} $-2\kappa \leq \bar{\kappa} \leq 0$. In the present context, these inequalities (applied to both the domain and the surrounding membrane) imply that the Gaussian modulus contrast must satisfy the bounds
\begin{equation}
    - 2 \kappa_d \leq \dg \leq 2 \kappa.
\end{equation}
With these constraints, any of the four universal regimes of curvature preference in Fig.~\ref{fig:universal} are accessible in principle. Experimentally, however, it is often observed that $\bar\kappa \approx -\kappa$ \cite{hu2012gaussian,deserno2015fluid}. Again applied to both the domain and the surrounding membrane, this suggests that we should typically expect $\dg \approx -\dk$, i.e.~the Gaussian modulus and bending rigidity contrasts will be of opposite sign and similar order of magnitude. With this  constraint, only regimes I, II, and IV of Fig.~\ref{fig:universal} would be accessible, with regime III remaining out of reach.

To be more concrete, let us consider model membranes with coexisting liquid-disordered ($L_d$) and liquid-ordered ($L_o$) phases as a potential experimental realization. The $L_d$ phase is generally substantially softer than the $L_o$ phase,
with bending rigidities typically of order tens of $k_{\mathrm B}T$ for $L_d$ and values reaching hundreds of $k_{\mathrm B}T$ for the $L_o$ phase, depending on composition and measurement method. \cite{baumgart2005elasticity,usery2017line}. Taking $20\,k_{\mathrm B}T$ and $100\,k_{\mathrm B}T$ as representative values, we would expect $\dk \approx \pm 80\,k_{\mathrm B}T$ and $\dg \approx \mp 80\,k_{\mathrm B}T$, where the upper and lower signs correspond to small $L_o$ domains in a $L_d$ matrix ($L_o$ in $L_d$), or to small $L_d$ domains in a $L_o$ matrix ($L_d$ in $L_o$), respectively. Line tensions between coexisting $L_o$ and $L_d$ domains are typically in the sub-pN to few-pN range, making $\lambda\approx 1\,\mathrm{pN} \approx 0.25\,k_{\mathrm B}T/\mathrm{nm}$ a representative value \cite{baumgart2003imaging,usery2017line}.

The line tension and the bending rigidity contrast (together with the constraint $\dg=-\dk$) define a characteristic domain radius $r_* = 4|\dk|/\lambda$, see Eq.~\ref{eq:critdomain}. Interestingly, for the estimates above, we obtain $r_* \approx 1.3\,\mu$m, which is in the typical range of domain sizes achievable in $L_o/L_d$ experiments with model membranes. In particular, Fig.~\ref{fig:universal} predicts that domains larger than this characteristic radius are found in regime II. Domains smaller than this critical radius, on the other hand, will be in regime I in the case of $L_o$ in $L_d$ (as $\dk>0$ in this case), while they will be in regime IV in the case of $L_d$ in $L_o$ (as $\dk<0$ in this case).

For these same parameters, we can rewrite the definition of $\hg$ in Eq.~\ref{eq:shapeparameters} as $\hg = \mp 1-r/r_*$, where the minus and plus signs correspond to $L_o$ in $L_d$ ($\dk >0$) and $L_d$ in $L_o$ ($\dk<0$), respectively. Comparing the accessible ranges as the domain radius increases from $r=0$ with the diagrams in Fig.~\ref{fig:shapes}, it appears that in the special case of zero spontaneous curvature ($m=m_d=\hm=0$) experiments would be able to test the transition equator-pole transition only in the case of $L_o$ in $L_d$ ($\dk >0$), as in the case of $L_d$ in $L_o$ ($\dk<0$) even small domains are already already have the same preferred configuration as large domains.

The magnitude of spontaneous curvature depends on the extent to which the membrane is asymmetric, and the cause of this asymmetry. For asymmetric exposure of small molecules such as sugars to the membrane, values as low as $m \approx 0.01\,\mu$m$^{-1}$ have been reported, whereas for asymmetric adsorption of BAR domain proteins, values as large as $m \approx 50\,\mu$m$^{-1}$ are possible \cite{10.1039/c2fd20105d}. Taking $m \approx m_d \approx 1\,\mu$m$^{-1}$ as a typical value, and $L = 10 \,\mu$m as a characteristic lengthscale of the curved membrane, we see from Eq.~\ref{eq:shapeparameters} that we can expect values of $|\hm| \approx |m|L \approx 10$, well within the range that allows exploration of the different behaviors predicted in the diagrams in Fig.~\ref{fig:shapes}. In fact, in order to explore values of $\hm$ of order one it is likely more convenient to use smaller spontaneous curvatures, such as those induced by asymmetric exposure to small molecules such as sugars and salts.

The above considerations show that experimentally accessible parameters could probe the different bifurcations that we predict regarding the spontaneous migration and preferred localization of small domains on the membrane. Importantly, however, lipid membranes are thermal systems subject to thermal fluctuations. Thus, it is important to ensure that the typical energy differences that cause domain migration are large relative to the thermal energy, $k_{\mathrm B}T$. Indeed, from Boltzmann statistics we expect that the ratio of probabilities of finding a domain at a point $A$ versus at a point  $B$ is given by $p_A/p_B = \exp{[(E_B-E_A)/(k_{\mathrm B}T)]}$, and energy differences $|E_B-E_A|\ll k_{\mathrm B}T$ can be washed out by thermal fluctuations. From Eq.~\ref{eq:central}, we have 
\begin{equation}
E_B-E_A = A\left[2\dk (M^2_B-M_A^2)
-4 \dm (M_B-M_A)
+\left(\dg-\frac{\lambda r}{4}\right)(K_B-K_A)\right]
\label{eq:centraldiff}
\end{equation}
Let us focus on the case of zero spontaneous curvature for simplicity, $\dm=0$. Additionally, for the typical parameters and domain sizes described above, with $|\dk|\sim |\dg|\sim \lambda r$, we see that the typical energies are of order $A|\dk(M^2_B-M^2_A)|$ and $A|\dk(K_B-K_A)|$, with $A=\pi r^2$ the area of the domain. Let us take the points $A$ and $B$ to be the equator and the poles. For a strongly prolate shape, with $c \gg a$, we have that $M_\mathrm{p}^2-M_\mathrm{e}^2 \approx K_B-K_A \approx c^2/a^4$. Using as an example $a=10 \,\mu$m, $c=50 \,\mu$m, $r=1 \mu m$, $|\dk|=80\,k_{\mathrm B}T$, we find energies with a typical scale of $\pi r^2 |\dk|c^2/a^4 \approx 60 k_BT$. Thus, the energy differences involved are significantly stronger than thermal fluctuations, and we expect the predicted effects to be observable in appropriately designed experiments.

\subsection{Conclusion}

We have shown that a small mobile lipid domain on a fixed curved membrane experiences a local energy landscape whose shape is controlled by the bending rigidity contrast, spontaneous curvature contrast, Gaussian modulus contrast, line tension and domain size.  This local energy yields a very simple universal classification for the curvature-preference of small domains,  whose regime boundaries are independent of the spontaneous curvature contrast. Further, the local energy  prediction of its preferred localization (or coexisting localizations) on any chosen shape, to leading order in the small size of the domain. Calculations for prolate and oblate spheroids show that even very simple closed surfaces can display both continuous and discontinuous migration, and that spontaneous-curvature contrast remains an essential control parameter in this case. Typically, we predict relocation of the domains upon growth by coalescence beyond a critical size, and comparison with experimental values suggests this to be a testable prediction.

Our work has various limitations, of which the most salient one is that it is restricted to the case where the membrane shape is fixed, e.g. by strong adhesion to a solid substrate, cortex, or scaffold.  If adhesion or scaffolding does not suppress local shape change, the domain can alter the curvatures that attracted it, and line tension can drive budding \cite{julicher1996shape}. Of course, a second important limitation is that our results are constrained to the regime in which the domains are much smaller than the local curvature radii of the membrane, as well as the characteristic lengthscale over which this curvature varies. As the domains grow beyond the realm of applicability of our approximation, the determination of preferred shapes and locations of membrane domains becomes a rather involved exercise \cite{fonda2018interface,fonda2019thermodynamic}. Finally, our work is based on a quasistatic approach, focused on determining the energy landscape experienced by the domain. Understanding the dynamics of domain migration will require dealing with the effects of membrane hydrodynamics and substrate friction, for which classical membrane-inclusion diffusion provides a starting point \cite{saffman1975}. These are left as avenues for future work.

\section*{Acknowledgments}
I thank Piermarco Fonda for discussions on the isoperimetric problem. Naturally, I would also like to acknowledge Wolfgang Helfrich, in whose honor this Special Issue is dedicated. His seminal 1973 paper \cite{helfrich1973} on the elasticity of lipid bilayers underlies much of the framework used not only in this work but throughout my entire career. I had the privilege of attending the Biomembrane Days meeting in Potsdam in 2012 celebrating his 80th birthday, only one month after starting my PhD. At the time, I did not yet know enough to appreciate what an extraordinary opportunity it was to meet Helfrich and so many of his former collaborators. I certainly do now.
\section*{Competing interests}
The author declares no competing interests.
\section*{Data and code availability}
All results are theoretical and contained within the paper.

\FloatBarrier
\printbibliography[title={References}]
\end{document}